\documentclass[%
    twocolumn,
    superscriptaddress,
    bibnotes,
    amsmath, amssymb, 
    aps, 
    prx,
    floatfix
]{revtex4-2}

\usepackage{hyperref}
\usepackage{graphicx}
\usepackage{braket}
\usepackage{mathtools}
 
\newcommand{\CCQ}{Center for Computational Quantum Physics, Flatiron Institute, New York, NY 10010, USA}
\newcommand{\CCM}{Center for Computational Mathematics, Flatiron Institute, New York, NY 10010, USA}

\newcommand{\smallsection}[1]{{{\it #1.}}}

\begin{document}

\title{Fast Tensor Network Imaginary Time Evolution by Implicit Stepping on \\ Logarithmic Grids}

\author{John P. Zima}
\affiliation{Department of Chemistry, Northwestern University, Evanston, Illinois 60208, USA}

\author{E. Miles Stoudenmire}
\affiliation{\CCQ}

\author{Steven R. White}
\affiliation{Department of Physics and Astronomy, University of California, Irvine, California 92697, USA}

\author{Olivier Parcollet}
\affiliation{\CCQ}
\affiliation{Universit\'e Paris-Saclay, CNRS, CEA, Institut de Physique Th\'eorique, 91191, Gif-sur-Yvette, France}

\author{Jason Kaye}
\affiliation{\CCQ}
\affiliation{\CCM}

\begin{abstract}

   We present a new method for the efficient imaginary time evolution of
   quantum many-body wavefunctions represented by matrix product states (MPS). We first show that logarithmic time grids are
   sufficient to resolve long imaginary time dynamics, yielding an exponential
   reduction in the number of time steps compared with standard approaches.
   We then show that A-stable implicit time-stepping methods for ordinary differential equations
   allow stable propagation for any time step size. The
   resulting scheme requires only matrix-vector products and
   linear solves, standard operations in the MPS toolbox. 
   We validate our approach with two examples: a Heisenberg spin
   chain, which we use to demonstrate a speedup of several orders of magnitude over the standard time-dependent
   variational principle method with uniform time steps, and a
   single-site Anderson impurity model with a metallic bath, for which
   propagation to large imaginary times allows one to observe the exponential
   dependence of the Kondo temperature on the interaction strength.

\end{abstract}

\maketitle

\smallsection{Introduction} 
Imaginary time dynamics---the evolution of a system in imaginary time $\tau$ 
from 0 to the inverse temperature $\beta$---plays a central role
in the theory of quantum many-body systems and in statistical quantum field theory.
Computing in imaginary rather than real time strikes a favorable balance
between computational cost and physical insight, 
as imaginary time correlation functions are typically much easier to compute than their real time counterparts~\cite{bloch1958undeveloppement,abrikosov1964methods}.
For example, tensor networks (TN) \cite{schollwock2011density}
evolving in imaginary time do not suffer from the entanglement growth driving up the cost of real time evolution \cite{stoudenmire2010minimally}, 
and quantum Monte Carlo methods operate almost exclusively in imaginary time to minimize the fermionic sign problem.
Yet imaginary time correlation functions encode a remarkably rich amount of
physical information: thermodynamic properties, the nature of the ground or
thermal state, and even low-energy dynamics.
%~\cite{feynman1953atomic,sachdev1993gapless,parcollet1999nflmott,werner2008spinfreezing,linden2020imaginary,dumitrescu2022planckian}.
Computing real time dynamics remains necessary to obtain certain
properties, such as transport coefficients or high-energy spectral functions, 
though some of these can be accessed via analytical continuation of imaginary time data \cite{GubernatisJarrell1996PhR}.

In TN methods, real and imaginary time evolution are typically implemented with the same algorithms. The most popular methods are probably time-evolving block decimation (TEBD)
\cite{vidal2004efficient} and the time-dependent variational principle (TDVP)
\cite{haegeman2011time} methods. The TDVP method is particularly efficient in many contexts, and is now widely favored, with uniform time steps tied to the energy scale of the Hamiltonian ~\cite{schollwock2011density,paeckel2019time,haegeman2016unifying}. In particular, the decaying nature of the $e^{-\hat{H} \tau}$ operator is not exploited. For obtaining thermodynamic properties at finite temperature, both the purification technique~\cite{feiguin2005finite} and minimally entangled typical thermal states (METTS)~\cite{white2009minimally,stoudenmire2010minimally} involve using imaginary time evolution to project an initial state towards the ground state, with the ground state recovered at long times. These methods become very expensive for very low temperature calculations due primarily to the  large number of time steps needed. 
For this reason, while TN methods have long offered a robust, accurate, and non-perturbative approach for quantum many-body simulations, 
they have not always been competitive with alternative methods (e.g.~perturbative and/or Monte Carlo-based) for imaginary time calculations due to large overhead costs.

In this paper, we revisit imaginary time evolution within the TN context, 
achieving a significant speedup and simplification using four key ideas.
First, contrary to
the conventional wisdom that step sizes are limited by the spectral radius of the
Hamiltonian or by a Trotter approximation, the imaginary-time-evolved wavefunction $\ket{\psi(\tau)}$ can be
resolved with exponentially growing time steps. Indeed, since $\ket{\psi(\tau)}$ can be expanded as a sum of decaying
exponentials $\exp(-E_i \tau)$,
the effective energy scale decreases with time and the characteristic time scale increases accordingly.
Second, we employ \textit{implicit A-stable} time stepping methods for ordinary differential equations (ODEs)
(using mathematical terminology, defined below), which have the property that propagation on arbitrarily large time steps is stable.
Third, these ideas can be applied efficiently in the high-dimensional space of matrix product states (MPS), since they rely on only two
operations, applying operators and solving linear systems, which are well-studied within the MPS framework~\cite{oseledets2012solution,Holtz:2012,dolgov2014alternating,bucci2024sketchgmres, pati1999dynamical, kuhner1999dynamical, Jeckelmann:2002}. 
And fourth, our ODE framework is flexible, supporting different time-stepping schemes, including high-order accurate methods which deliver exceptional computational efficiency.
This contrasts with TEBD and TDVP, which both introduce a typically low-order operator splitting error~\cite{haegeman2016unifying}.
Overall, our approach lowers the cost of tensor-network-based imaginary time evolution enough to significantly broaden their applicability in a variety of contexts, several of which are discussed in the conclusion. 

We demonstrate the efficiency of our approach on 
a Heisenberg spin chain, comparing with a standard TDVP implementation. We then consider the Anderson quantum impurity model,
recovering the Kondo scaling of the double occupancy in the regime of
exponentially small Kondo temperature. This is a challenging problem, requiring propagation to very large imaginary times. 
Combining our time evolution method with the METTS algorithm yields the result with modest computational cost.

\smallsection{Method}
The imaginary time evolution of a state $\ket{\psi}$ under the Hamiltonian $\hat{H}$, which we assume to be positive semi-definite and finite-dimensional,  
is given by $\ket{\psi(\tau)} = e^{-\hat{H}\tau} \ket{\psi(0)}$.
By a shift, we assume that the ground state energy is 0, and we denote the largest eigenvalue of $\hat{H}$ by $E_\text{max}$.
We will represent the state $\ket{\psi(\tau)}$ by an MPS and $\hat{H}$ by a matrix product operator (MPO)~\cite{hubig2017generic,schollwock2011density}.
The solution formula implies that each component of
$\ket{\psi(\tau)}$ is a linear combination of decaying exponentials 
with decay rates given by the eigenvalues $E_i$ and coefficients by the eigenvectors $\ket{v_i}$ of $\hat{H}$:
\begin{equation} \label{eq:psi_decomp}
   \ket{\psi(\tau)} = \sum_{i} e^{ -E_i \tau} \ket{v_i} \braket{v_i |\psi(0)}.
\end{equation}

It can be shown rigorously that a logarithmic grid of $N$ nodes clustered towards
the origin $\tau = 0$, with smallest spacing 
$\Delta \tau_\text{min} = \mathcal{O}(E_\text{max}^{-1})$, 
{\it uniformly} resolves any set of exponentials
 $\{e^{-E \tau}\}_{0 \leq E \leq E_\text{max}}$ to a given
tolerance \cite{gimbutas20, kaye22_dlr}. 
It is therefore sufficient to obtain $\ket{\psi(\tau)}$ on such a grid, which 
contains $N = \mathcal{O}(\log(E_\text{max} \tau_{\max}))$ nodes. 
By contrast, using uniform time steps
requires $\Delta \tau = \mathcal{O}(E_\text{max}^{-1})$, or $N = \mathcal{O}(E_\text{max} \tau_{\max})$. 
We note that this is the main idea underlying the compact discrete Lehmann
representation of imaginary time Green's functions \cite{kaye22_dlr}, but
applied to the wavefunction itself.
We note two previous works exploring ideas similar to logarithmic stepping in the context of TN imaginary time evolution,
both focusing on the finite-temperature density operator rather than a generic state vector as in this work: Ref.~\onlinecite{chen18} performs repeated self-multiplication of an MPO approximation of $e^{-\tau \hat{H}}$ at small $\tau$, while Ref.~\onlinecite{li23} evolves the density operator using a TDVP-like method on a grid which is initially logarithmic, but eventually uniform at large times. 

We view $\ket{\psi(\tau)}$ as the solution of the linear ODE 
\begin{equation} \label{eq:imtime_schrod}
   \frac{d}{d\tau} \ket{\psi(\tau)} = -\hat{H} \ket{\psi(\tau)}
\end{equation}
with given initial data. 
Integrating \eqref{eq:imtime_schrod} between grid points $\tau_{k-1}$ and $\tau_k$ yields 
\begin{equation} \label{eq:ode_integral}
    \ket{\psi(\tau_k)} = \ket{\psi(\tau_{k-1})} - \hat{H} \int_{\tau_{k-1}}^{\tau_k} \ket{\psi(s)} ds,
\end{equation}
and approximating the integral by the trapezoid rule yields the
implicit trapezoid rule \cite{iserles1996first,jose2006time} time stepper
\begin{equation}
\label{eq:trapezoid}
\left(I+\frac{\Delta \tau_k}{2} \hat{H}\right)
\ket{\psi^{(k)}} =
\left(I-\frac{\Delta \tau_k}{2} \hat{H}\right)
\ket{\psi^{(k-1)}} ,
\end{equation} 
where $\ket{\psi^{(k)}}$ approximates
$\ket{\psi(\tau_k)}$ and $\Delta \tau_k = \tau_k - \tau_{k-1}$. Here, $\ket{\psi^{(k)}}$ is \textit{implicitly} defined,
and must be obtained by solving a linear system at each time step.
 
A key insight is that \eqref{eq:trapezoid} can be solved with the logarithmic
grid $\{\tau_k\}$ discussed above.  Indeed, as a consequence of the previous
discussion, such a grid is sufficient to resolve $\ket{\psi(\tau)}$, but one
might wonder whether a given time-stepping method is accurate and stable on a
grid with growing time steps. \textit{Explicit} ODE methods like the forward
Euler method, which involve only matrix-vector products and no linear solve, are not: crossing a threshold step size causes the discrete time
stepper to become \textit{unstable}, artificially amplifying high-energy modes and
leading to eventual blow-up.
However, the implicit trapezoid rule is \textit{A-stable}:
there is no restriction on the time step size for stability, and the accuracy of the approximate solution is
determined only by the accuracy with which $\ket{\psi(\tau)}$ is resolved. In particular, exponentially-growing time steps are
permissible. 
No explicit method has this property, and therefore only A-stable, implicit ODE solvers are compatible
with logarithmic grids for arbitrarily large propagation times.
The accuracy and stability properties of the implicit
trapezoid rule, as contrasted with explicit methods, are discussed further in
Appendix~\ref{sec:trapezoidrule}, along with the definition of A-stability.

In practice, instead of precisely logarithmic time grids, we use $P$ panels
with endpoints
$\{0, \tau_\text{max}/2^{P-j} \text{ for } j\in [1,P]\}$
and
$n$ uniform time steps on each panel. This approach yields an
$\mathcal{O}(n^{-2})$ order of convergence for the trapezoid rule.

To improve the efficiency of the scheme, we also explore a scheme which replaces the trapezoid
approximation with an implicit, A-stable, and \textit{spectrally accurate} Gauss-Legendre collocation approach.
The solution is approximated in time by a Legendre polynomial expansion, rather than a uniform grid, on each
panel, and the coefficients of the expansion are obtained by enforcing the integral equation (\ref{eq:ode_integral}) 
(with limits of integration replaced by panel endpoints) at the Gauss-Legendre nodes. 
This leads to a single linear system for the Legendre coefficients on each panel.
In what follows, we mainly focus on the trapezoid rule for conceptual simplicity, but demonstrate the performance of the spectral method in the next section. The details of this solver will be described in an upcoming publication.

In the case of (\ref{eq:trapezoid}), advancing a time step requires performing two operations: (i) form the right hand side by applying $(I-\Delta \tau_k \, \hat{H}/2)$  
and (ii) solve \eqref{eq:trapezoid} as a linear equation for $\ket{\psi^{(k)}}$.
Both are well-studied operations in the MPS/MPO format~\cite{schollwock2011density,camano2026successive},
and in particular several MPO-MPS linear solvers are
described in the
literature~\cite{oseledets2012solution,Holtz:2012,dolgov2014alternating,bucci2024sketchgmres}. The required operations are the same for any implicit ODE-type solver. 
We use the ITensor software \cite{fishman2022itensor} to solve the
MPO-MPS linear systems, using a two-site DMRG-style scheme
with GMRES~\cite{saad1986gmres} local solves (see Appendix \ref{app:tn_details} for technical details).
Crucially, we observe for the systems studied that the cost of solving \eqref{eq:trapezoid} appears approximately independent
of the time step size $\Delta \tau_k$, i.e., solving with a fixed number of sweeps and local iterations yields
approximately uniform accuracy as the propagation time increases.

At fixed bond dimension $\chi$ of the MPS, this directly yields a logarithmic scaling $\mathcal{O}(\log(\tau_{\max}))$ of the total computational cost due to the use of logarithmic time grids.
This is to be contrasted with the $\mathcal{O}(\tau_{\max})$ scaling of standard methods such as TDVP based on uniform time-stepping~\cite{haegeman2011time,haegeman2016unifying,paeckel2019time,yang2020time,li2024time}. For cases
in which the bond dimension $\chi(\tau)$ is determined adaptively and allowed to grow, we analyze two realistic scenarios in Appendix \ref{sec:computational_costs}:
logarithmic growth, $\chi(\tau) = \log \tau$, and power law growth, $\chi(\tau) = \tau^\alpha$ for some $\alpha > 0$. In both cases, we find the reduction in cost to be proportional to $\tau_{\max}$ up to logarithmic factors, the same as for the fixed bond dimension case. 

We obtain a remarkably simple imaginary time evolution scheme which, assuming
the fixed linear solve cost observed in practice persists, yields a dramatic
speedup over common methods such as uniform-stepping TDVP by making use of the
particular structure of imaginary time dynamics. We also make an empirical
observation below that a standard implementation of TDVP using logarithmic time
steps demonstrates a similar instability to that of explicit ODE methods.
Furthermore, we favor our approach for its simplicity and modularity: unlike
for TEBD and TDVP, the main computational machinery---in our case an MPO-MPS
linear solver---is decoupled from the choice of time-stepping scheme, so that
advances in the rapidly developing field of MPO-MPS linear solvers can directly
benefit our method.

\smallsection{Heisenberg chain}
As a first test of our method, we consider a one-dimensional chain of $L=100$ quantum $S=1/2$ spins governed by the antiferromagnetic Heisenberg Hamiltonian
\[
\hat{H} = J \sum_{j=1}^{L-1} \vec{S}_{j} \cdot \vec{S}_{j+1},
\]
with open boundary conditions, where we use the antiferromagnetic coupling $J=1$.

We perform an imaginary time evolution from the Néel state, a type of calculation that is a building block 
for more elaborate techniques like METTS~\cite{white2009minimally,stoudenmire2010minimally}.
\begin{figure}[t] \includegraphics[width=\columnwidth]{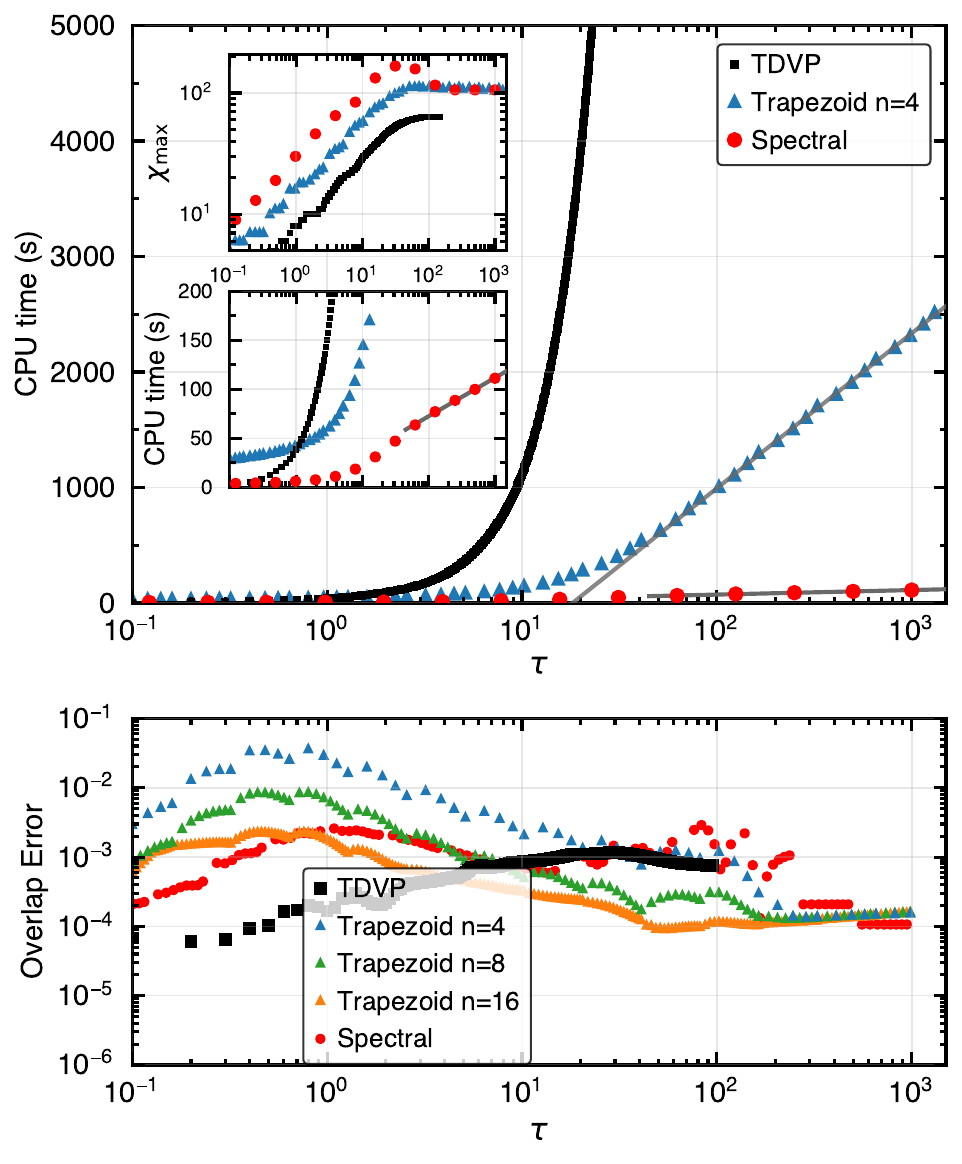}
   \caption{Imaginary time evolution, starting from a N\'{e}el
      state, for a 100-spin Heisenberg chain. 
      Top panel: CPU time vs. $\tau$ for uniform-stepping TDVP with a time step of 0.1, trapezoid method with $n=4$ time steps per panel, and spectral 
      method with comparable accuracy. Upper inset: bond dimensions of methods shown (taking max over all bonds), fixing the discarded weight 
      (i.e., total discarded squared Schmidt weight at a bond) to $10^{-10}$. 
      Lower inset: zoom of CPU times. The gray lines are linear
      fits of the last decade in $\tau$, showing logarithmic scaling with
      respect to $\tau$ of the trapezoid and spectral methods once the bond dimension saturates.  
      Bottom panel: Overlap error
      $\sqrt{1-\braket{\psi|\psi_\mathrm{ref}}}$ (for normalized states) compared to well-converged reference
      results $\ket{\psi_\mathrm{ref}}$. Convergence of the trapezoid method is second-order with respect to $n$. 
\label{fig:HeisenbergChainMainResult} } \end{figure}
Fig.~\ref{fig:HeisenbergChainMainResult} shows the CPU time (single core) of various methods
vs. $\tau$ for an adaptively-determined bond dimension $\chi$ (top panel), and 
the overlap error at each time step relative to a well-converged
reference result (bottom panel).

Using the trapezoid method, we vary the number of time steps $n$ per panel, 
demonstrating second-order convergence (doubling $n$ improves accuracy by a factor of about 4).
We observe logarithmic-scaling CPU times (grey lines), 
as well as linear scaling for the standard uniform time step TDVP method. 
The initial faster-than-logarithmic scaling is due to growth of the bond dimension $\chi$,
as shown in the inset, and an $\mathcal{O}(\chi^3)$ cost per time.
The bond dimension only slightly overshoots the long-time,
ground-state value.

The comparison of TDVP with the trapezoid rule clearly demonstrates the scaling
advantage of the latter, but for further efficiency we also use the spectral method described above. This further improves the prefactor in
the computational cost of our approach: the top panel of
Fig.~\ref{fig:HeisenbergChainMainResult} shows an almost negligible cost of this
spectral method compared with TDVP, with the inset showing that the full time evolution is completed in approximately 100 seconds.

\begin{figure}[t] 
   \includegraphics[width=\columnwidth]{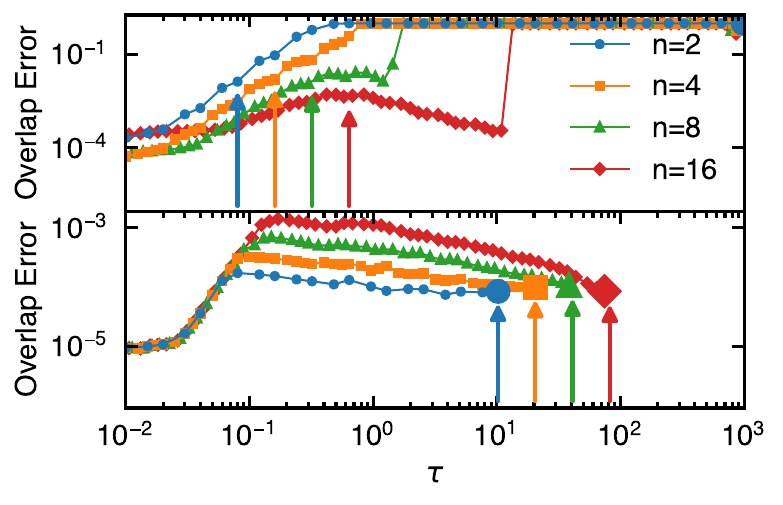}
   \caption{
(Top) Overlap errors $\sqrt{1-\braket{\psi|\psi_\mathrm{ref}}}$ (for normalized states) 
for the Heisenberg chain. Same parameters as in Fig.~\ref{fig:HeisenbergChainMainResult}, 
except that we use an {\it explicit} time stepper, Heun's method (cf text).
The vertical arrows indicate the critical time step at which blow-up is theoretically expected to begin, and the trajectory becomes unstable
shortly after. (Bottom)~Overlap error using TDVP with a local
Krylov solver on a logarithmic grid. 
Arrows indicate a halting of the TDVP solver due to blow-up in the local solves. 
 \label{fig:Heun}
}
\end{figure}

To demonstrate the necessity of using A-stable implicit methods, the top panel of Fig.~\ref{fig:Heun} shows
the result of replacing the implicit trapezoid rule \eqref{eq:trapezoid} with an explicit time-stepping scheme, which has the advantage of avoiding linear solves. We use Heun's method, a second-order explicit predictor-corrector version of the trapezoid rule, and by computing the stability region of the method and the largest eigenvalue of the $\hat{H}$ (see Appendix~\ref{sec:trapezoidrule}), we determine the critical time step size $\Delta \tau$ at which the method begins to amplify high-energy components of the numerical solution (vertical arrows). We indeed observe instability after this critical time step size is exceeded (note that doubling $n$ doubles the time $\tau$ at which the critical step size is reached on our logarithmic grid). We have empirically observed a similar instability using TDVP with a Krylov-based local solver on a logarithmic grid: the bottom panel of Fig.~\ref{fig:Heun} shows eventual failure of the TDVP time step due to an instability in the local solves, with the failure time again proportional to $n$.

\smallsection{Anderson impurity model}
\begin{figure}[b]
\centering
\includegraphics[width=\columnwidth]{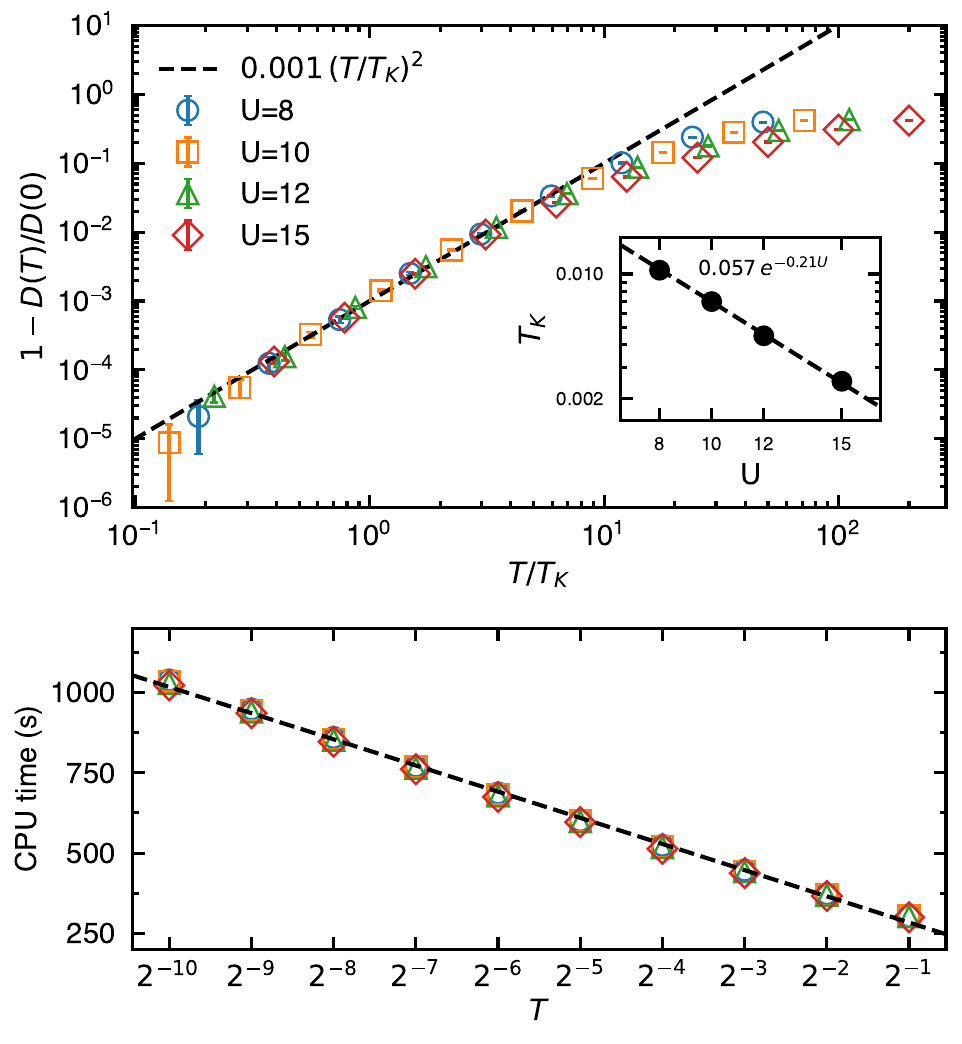}
\caption{\label{fig:collapse} 
   (Top) Double occupancy $D(T)$ 
   of the Anderson impurity model vs. $T/T_K(U)$. $T_K(U)$ are adjusted to obtain a collapse 
   of the curves for various $U$. Error bars are 95\% confidence intervals of
   METTS sampling error.  $D(0)$ is measured directly in the ground state.
   (Inset) $T_K$ vs. the interaction strength $U$, showing an exponential dependence  on $U$. 
   The dashed curve is a fit $T_K(U) \approx 0.057 \times e^{-0.21 U}$.
   (Bottom) Average CPU time of the computation of individual METTS versus temperature for different $U$, 
       showing a $\log \beta$ scaling. The straight dashed line is a linear fit.
}
\end{figure}
We next consider a more challenging case:
the Anderson impurity model,
defined by the Hamiltonian 
\begin{equation*}
\hat H =
\epsilon_d n_d
+ U n_{d\uparrow}n_{d\downarrow}
+ \sum_{\substack{1\le j\le N_B\\ \sigma=\uparrow,\downarrow}}
   \epsilon_j \hat c^\dagger_{j\sigma}\hat c_{j\sigma}
   +
   \left(V_j \hat d^\dagger_\sigma \hat c_{j\sigma} + h.c.\right)
\end{equation*}
where $\hat{d}_\sigma$ is the annihilation operator for the impurity and
$\hat{c}_{j\sigma}$ for bath site $j$, $n_{d\sigma} \equiv  \hat{d}^\dagger_{\sigma} \hat{d}_{\sigma}$, 
and $n_d \equiv n_{d\uparrow} + n_{d\downarrow}$. We use $\epsilon_d = -U/2$, and bath coupling $V_j$ and energies $\epsilon_j$ obtained from a semi-circular density of states, as detailed in 
Appendix~\ref{sec:aim}. 
We use an efficient sum-of-poles approximation to achieve high accuracy with $\mathcal{O}(\log
\beta)$ bath sites \cite{huang25,huang23}, where $\beta$ is the largest inverse temperature considered.

The Anderson model exhibits a parametrically exponentially-low energy scale---the Kondo temperature $T_K(U)$. 
It therefore requires exponentially long imaginary-time evolution at low temperatures.
This is particularly challenging for methods using uniform time steps, but
exponentially low temperatures can be reached in linear computing time at fixed bond dimension using our approach.

We focus on the behavior of the double occupancy
$D(T) = \langle n_{d\uparrow}n_{d\downarrow}\rangle$
as a function of $T$.
To compute at finite temperature, we use the METTS
algorithm~\cite{white2009minimally,stoudenmire2010minimally}, which samples
finite-temperature expectation values by generating a Markov chain of
normalized imaginary-time-evolved product states. 
For each combination of $U \in \{8,10,12,15\}$ and $\beta = 2^k$, $k=1,\ldots,10$, we used a constant bond
dimension $\chi = 200$, sufficient to represent the ground state to high accuracy, and used the trapezoid rule time stepper.
The number of METTS sampled varied from about $50,000$ at the highest temperature $\beta=2$ down to $15,000$ at the lowest temperature $\beta=1024$.
Further implementation details are provided in Appendix ~\ref{sec:aim}.
At low $T$, the double occupancy behaves as \cite{dirks2013double} 
\begin{equation} \label{eq:double_occ_asymptotics}
D(T) = D(0)(1-a (T/T_K(U))^2 + \ldots)
\end{equation}
where $a$ is a constant and the Kondo temperature $T_K(U)$ decays exponentially with $U$ at large $U$.
Our results are presented in Fig.~\ref{fig:collapse}, which shows $1-D(T)/D(0)$ versus $T$ (top panel).
The curves for various $U$ collapse on a universal scaling curve.
This scaling behaviour allows us to extract $T_K(U)$, shown in  Fig.~\ref{fig:collapse} (inset).
We check that $T_K(U)$ decays exponentially with $U$.
Our results are compatible with those obtained using the numerical renormalization group
(NRG)~\cite{wilson1975renormalization} on a similar model~\cite{dirks2013double}.

The key result is presented on the second panel of Fig.~\ref{fig:collapse}, 
which shows the average CPU time for a full METTS loop (one imaginary
time evolution) against temperature.
We observe logarithmic scaling with $\beta$ over a large range of temperatures $\beta \in[ 2, 1024]$,
allowing us to reach very low temperatures at a modest cost.

\smallsection{Conclusion}
We have demonstrated that combining stable, implicit ODE solvers on logarithmic
time grids with tensor network methods yields a dramatic reduction in
computational cost for imaginary time evolution, compared with
standard methods based on uniform time stepping. We applied our method to the Heisenberg
chain and the Anderson model, for which the use of
logarithmic time grids allows us to resolve the exponentially low Kondo temperature. 

Imaginary time evolution is ubiquitous in computational quantum many-body physics, so significantly lowering the cost of high-precision imaginary time evolution has the potential to make the TN approach newly competitive in a variety of applications. These include the preparation of steady states in open quantum system simulations via the 
Lindblad master equation~\cite{zwolak2004mixed}, the calculation of imaginary time Green's functions, 
and other methods requiring imaginary time evolution at large propagation 
times, such as METTS~\cite{wietek2021stripes,wietek2021mott,feng2023bose,sinha2025forestalled,
wang2026spectroscopy,wang2026anomalous,chalopin2026observation,sinha2026evolution}, purification~\cite{feiguin2005finite,schollwock2011density,
verstraete2004matrix,bauernfeind2022minimally}, and MPS thermal pure 
states~\cite{iwaki2021thermal}.
In dynamical mean-field theory for strongly correlated systems~\cite{georges96,
kotliarRMP2006}, combining our method with efficient bath discretization 
tools~\cite{mejuto2020,huang23,huang25,kaye22_dlr} provides a 
path towards a general non-perturbative impurity solver whose computational cost scales as a low power of $\log \beta$. 
Quantum Monte Carlo methods~\cite{Rubtsov05,werner06a,werner06b,Gull:2008ab,gull11} 
are standard solvers in this field, 
but have several limitations, such as a slow convergence, poor scaling with temperature, and the fermionic sign problem.
Fast imaginary time TN solvers open a path to overcoming all three of these issues and resolving low energy scales in correlated metals
currently out of reach.  

Beyond quantum physics, our approach could be applied to classical master 
equations relaxing to a steady state, by representing the time-dependent 
probability distribution as a tensor network. 
Such tensor network simulations 
have already been employed in disease spreading 
dynamics~\cite{Dolgov2024}, reaction-rate dynamics in 
chemistry~\cite{Nicholson2023Quantifying,Zima2025Chemical,liao2015tensor}, 
and CMOS electronic devices~\cite{murphy2026dissipation}. In these settings, 
long-time evolution is essential for capturing phenomena with long time scales, 
such as rare switching events between metastable basins in configuration 
space~\cite{Zima2025Chemical}.

We lastly note that our approach requires only a compressed state representation 
supporting operator applications and linear solves. Beyond tensor networks, it 
could therefore potentially be applied to imaginary time evolution of neural network 
quantum states~\cite{Carleo_2017,hendry22,vivas2022neural,ledinauskas2023scalable,vandewalle25,sinibaldi26}.

\acknowledgments

JPZ is grateful for support from the Flatiron Institute during the Pre-Doctoral Researcher program. SRW was supported by the U.S. NSF under Grant
DMR‑2412638. The Flatiron Institute is a division of the Simons Foundation. 

\appendix

\section{Accuracy and stability of the trapezoid rule approximation} \label{sec:trapezoidrule}

We begin by deriving the trapezoid rule approximation of the imaginary time dynamics. Given the state at time $\tau$, we can use \eqref{eq:imtime_schrod} to write the state at time $\tau + \Delta \tau$ as
\begin{equation} \label{eq:ode_integral_step}
    \ket{\psi(\tau + \Delta \tau)} = \ket{\psi(\tau)} - \hat{H} \int_\tau^{\tau + \Delta \tau} \ket{\psi(s)} ds.
\end{equation}
The trapezoid rule is obtained by making the local trapezoidal approximation to the integral, namely
\begin{equation}
    \ket{\psi(\tau + \Delta \tau)} \approx \ket{\psi(\tau)} - \frac{\Delta \tau}{2} \hat{H} (\ket{\psi(\tau)} + \ket{\psi(\tau + \Delta \tau)})
\end{equation}
which gives \eqref{eq:trapezoid}. 

Let us now consider the error of the trapezoid rule approximation of the integral for the time steps in our logarithmic grid. Rather than a complete error analysis of our time stepping scheme, we give a simple argument justifying this grid. In short, for large panels, modes with a rapid decay rate are smaller and flatter, so one can achieve the same accuracy with a larger time step. To see this, we consider the approximation error for each mode $e^{-\tau \lambda}$, where $\lambda$ is an eigenvalue of $\hat{H}$. For an arbitrary interval $[a, b]$ with $h = b-a$, we have
\begin{equation}
    \begin{aligned}
        E_{[a,b]}(\lambda) &= \frac{\lambda h}{2} (e^{-a \lambda} + e^{-b \lambda}) - \lambda \int_a^{b} e^{-\tau \lambda} d\tau \\ 
        &= \frac{\lambda h}{2} (e^{-a \lambda} + e^{-b \lambda}) - (e^{-a \lambda} - e^{-b \lambda}) \\
        &= e^{-\lambda (a+b)/2} \left( \mu \cosh(\mu/2) - 2\sinh(\mu/2) \right)
    \end{aligned}
\end{equation}
for $\mu = \lambda h$ the dimensionless product of the mode decay scale and panel width. 
We make the simplifying assumption $n = 1$ for our logarithmic grid, i.e., a single time step per interval, and $\tau_{\max}$ a power of 2, so that $a = h = 2^k$ and $b = 2^{k+1}$ for $k$ defining the panel. In this case, the error is a function of $\mu$ alone:
\begin{equation}
    E(\mu) = e^{-3 \mu /2} \left( \mu \cosh(\mu/2) - 2\sinh(\mu/2) \right).
\end{equation}
$E(\mu)$ is bounded for $\mu > 0$, achieving a maximum of approximately $0.04$ at $\mu \approx 2.1$. Thus the local error is uniform, independent of the panel index.

Even if a particular grid is sufficient to resolve the solution, a time stepping procedure might fail to produce stable discrete dynamics. The simplest example of this is the forward Euler method, an explicit scheme with first-order accuracy. It is obtained by making the left-endpoint approximation to the integral in \eqref{eq:ode_integral_step}:
\begin{equation}
    \ket{\psi^{(n+1)}} = \ket{\psi^{(n)}} - \Delta \tau \hat{H} \ket{\psi^{(n)}} = (I - \Delta \tau \hat{H}) \ket{\psi^{(n)}}.
\end{equation}
The discrete solution is simply
\begin{equation}
    \ket{\psi^{(n)}} = (I - \Delta \tau \hat{H})^n \ket{\psi_0}.
\end{equation}
Thus if $\hat{H}$ has an eigenvalue $\Lambda$ such that $|1 - \Delta \tau \Lambda| > 1$, the discrete solution will diverge. Thus we require $\kappa = \Delta \tau \Lambda \leq 2$ for stability, and a logarithmic grid cannot be used, even though it resolves the solution itself.

The \textit{stability region} of a time-stepping method is defined as the set of complex numbers $z = \lambda \Delta \tau$ such that the approximate solution of the ODE $y' = -\lambda y$ remains bounded (note that we have flipped the sign of $\lambda$ compared with convention). Thus, the stability region of the forward Euler method is the closed unit disk centered at $1$ in the complex plane. For Hermitian $\hat{H}$, we need only consider the stability region on the real line, yielding the stability condition above. A method is called A-stable if its stability region includes the entire right half-plane; in particular, for Hermitian $\hat{H}$ with positive eigenvalues, there is no stability restriction on the time step size. 
For the trapezoid rule, we have
\begin{equation}
    \ket{\psi^{(n+1)}} = \left((I + \frac{\Delta \tau}{2} \hat{H})^{-1} (I - \frac{\Delta \tau}{2} \hat{H})\right)^n \ket{\psi_0},
\end{equation}
and the stability condition is $|(1 - \frac{\Delta \tau}{2} \lambda) / (1 + \frac{\Delta \tau}{2} \lambda)| \leq 1$ for all eigenvalues $\lambda$ of $\hat{H}$. This condition is satisfied for any $z = \lambda \Delta \tau$ in the right half-plane, so the trapezoid rule is A-stable and has no stability restriction on the time step size: one can design the grid purely based on accuracy considerations. No explicit method can be A-stable (or includes the positive real axis in its stability region). Thus, our logarithmic grid approach involving growing time steps can be used with the implicit trapezoid rule, but not with explicit methods to arbitrarily large propagation times.

\section{Technical details of tensor network calculations}
\label{app:tn_details}

To generate the MPO tensor network form of the Hamiltonians we use, namely the Heisenberg spin chain and the single-impurity Anderson model, we use standard MPO compression techniques to sum and generate these operators as MPOs to machine precision with modest bond dimensions \cite{nakatani2016mpo, parker2020mpo}. The ITensor ``OpSum'' system was used to programatically generate all MPOs \cite{fishman2022itensor}.

The trapezoidal time stepping method and its spectral variant involve operator application and linear solving steps. To perform these, we use the \href{https://github.com/ITensor/ITensorMPS.jl}{ITensorMPS.jl} package which is part of the ITensor ecosystem \cite{fishman2022itensor}. The operator application is performed using the ``density matrix'' algorithm for MPO-MPS products \cite{ma2024approximate}. The linear solving step uses a DMRG-like algorithm but with the ``$b$'' MPS projected into the same basis as the ``$x$'' MPS (for solving an $Ax=b$ problem) and the GMRES algorithm as the solver in each basis configuration. To aid in convergence, the right-hand side $\ket{b^{(k)}} = (I-\frac{\Delta \tau_k}{2} \hat{H}) \ket{\psi^{(k-1)}}$, a forward Euler half-step, is used as an initial guess.

\section{Analysis of computational costs with varying bond dimension \label{sec:computational_costs}}

Consider a varying bond dimension $\chi(\tau)$, and either a uniform time grid $\tau_k = k$ or a logarithmic time grid $\tau_k = 2^k$ for $k=1,2,\ldots$, and a maximum time $\tau_{\max} = 2^m$. 
The cost of uniform time stepping is $C_\mathrm{U} = \sum_{k=1}^{2^m} \chi^3(k)$ and the cost of logarithmic time stepping is $C_\mathrm{L} = \sum_{k=1}^m \chi^3(2^k)$. 
For fixed $\chi$, we obtain $\mathcal{O}(\chi^3 \tau_{\max})$ and $\mathcal{O}(\chi^3 \log(\tau_{\max}))$ scaling, respectively. 
If the bond dimension grows logarithmically, $\chi(\tau) = \log \tau$, then $C_\mathrm{U} \sim m^3 2^m = \mathcal{O}(\tau_{\max} \log^3(\tau_{\max}))$ by a simple generalization of Stirling's formula, and $C_\mathrm{L} = \sum_{k=1}^m \log^3(2^k) \sim m^4 = \mathcal{O}(\log^4(\tau_{\max}))$, so we again find an exponential reduction in cost. 
For power law growth $\chi(\tau) = \tau^\alpha$, an integral approximation yields $C_\mathrm{U} \sim (2^m)^{3\alpha + 1} = \mathcal{O}(\tau_{\max}^{3\alpha + 1})$ and geometric series summation yields $C_\mathrm{L} \sim (2^m)^{3\alpha} = \mathcal{O}(\tau_{\max}^{3\alpha})$, so we find an $\mathcal{O}(\tau_{\max})$ reduction in cost. 
Up to a possible factor $\log(\tau_{\max})$, the cost reduction is proportional to $\mathcal{O}(\tau_{\max})$ in all cases.

\section{Details of the Anderson impurity model calculation} \label{sec:aim}

We use the METTS algorithm \cite{white2009minimally,stoudenmire2010minimally} to compute the thermal properties of the Anderson impurity model.
Beginning from a product state $|i\rangle$, one forms the state $|\psi_i\rangle \propto e^{-\beta \hat H/2}|i\rangle$ with $\beta = 1/T$, measures it in a product basis to obtain a new product state $|j\rangle$ with probability $|\langle j|\psi_i\rangle|^2$, and iterates this procedure. Thermal expectation values $\langle \hat{O}\rangle_\beta$ are estimated by taking an unweighted average over the METTS $|\psi_i\rangle$, namely $\langle \hat{O}\rangle_\beta \approx \frac{1}{N_\text{M}} \sum_{i=1}^{N_\text{M}} \langle \psi_i|\hat{O} |\psi_i\rangle $.
We seed each METTS Markov chain with a product state sampled from the ground state (computed using DMRG). 

We estimate the double occupancy by taking $\hat{O} = \hat{d}^\dagger_\uparrow \hat{d}_\uparrow \hat{d}^\dagger_\downarrow \hat{d}_\downarrow$. We perform the repeated time evolutions to obtain $|\psi_i\rangle$ using our logarithmic stepping scheme with the first panel $[0,0.01]$, and $n=4$ steps per panel. As in our Heisenberg model study, we use a single GMRES restart with a Krylov subspace of size $10$ for each local solve during the linear solve, as part of a single DMRG-style sweep. However, unlike in the case of the Heisenberg model study, here we not only normalize the state after each step, but also shift the operator's spectrum by the energy of the new time-evolved state, in order to minimize the norm drift of the next step. We use a maximum bond dimension of 200 for each MPS representing the state, and keep all non-zero singular values when the bond dimension is below the cap.

Our solver is compatible with efficient, $\mathcal{O}(\log \beta)$-scaling bath discretization algorithms \cite{mejuto2020,huang23,huang25,kaye22_dlr}. We use the sum-of-poles hybridization fitting scheme described in Refs.~\cite{huang23,huang25} implemented in the \texttt{adapol} package \cite{adapol_repo}, which combines the AAA rational approximation algorithm \cite{nakatsukasa18} with nonlinear optimization. Since we treat the same impurity model at multiple temperatures, it is necessary to generate a single bath discretization which is valid across the full temperature range $\beta \in [2,1024]$. To do so, we use \texttt{adapol} to generate a sum-of-poles approximation of the hybridization function $\Delta$ with semicircular spectral density $\rho(\omega) = 2/\pi \sqrt{1-\omega^2}$ which is accurate to a pointwise tolerance $\varepsilon = 10^{-6}$ for all imaginary frequencies $i \omega > i \pi/\beta_{\max}$, the smallest Matsubara frequency for the lowest temperature $\beta_{\max} = 1024$. Running \texttt{adapol} with $\beta = 1024$ is not sufficient to accomplish this, since the approximation is only guaranteed to be accurate at the Matsubara frequencies corresponding to the input $\beta$, and we have observed that the accuracy tolerance is violated between the lowest few Matsubara frequencies. Rather, we run \texttt{adapol} with a much larger value $\beta = 10000$, which has the effect of making the Matsubara sampling grid denser, and verify a posteriori that the resulting approximation satisfies the condition stated above. The resulting bath has 29 sites.
It would be useful to develop a scheme which directly generates a single approximation valid across a range of temperatures, rather than a single temperature, but we leave this to future work.

\bibliography{main}

\end{document}